\documentclass[]{pasj01}
\Received{$\langle$reception date$\rangle$}
\Accepted{$\langle$acception date$\rangle$}
\Published{$\langle$publication date$\rangle$}

\usepackage{bm}

\begin{document}

\title{Extracting common signal components from the X-ray and optical light curves of GX 339$-$4: new view for anti-correlation}
\author{Tomoki Omama\altaffilmark{1, 2, 3}
        Makoto Uemura\altaffilmark{4},
        Shiro Ikeda\altaffilmark{5},
        Mikio Morii\altaffilmark{6}}%
\altaffiltext{1}{Department of Physical Science, Hiroshima University, Kagamiyama 1-3-1, Higashi-Hiroshima 739-8526, Japan}
\altaffiltext{2}{Institute of Space and Astronautical Science, Japan Aerospace Exploration Agency, 3-1-1 Yoshonidai, Chuo-ku, Sagamihara, Kanagawa 252-5210, Japan}
\altaffiltext{3}{Department of Space and Astronautical Science, School of Physical Science, The Graduate University for Advanced Studies, SOKENDAI, 3-1-1 Yoshinodai, Chuo-ku, Sagamihara, Kanagawa 252-5210, Japan}
\altaffiltext{4}{Hiroshima Astrophysical Science Center, Hiroshima University, 1-3-1 Kagamiyama, Higashi-Hiroshima 739-8526, Japan}
\altaffiltext{5}{The Institute of Statistical Mathematics,10-3 Midori-cho, Tachikawa, Tokyo 190-8562, Japan}
\altaffiltext{6}{DATUM STUDIO CO., LTD., Toranomon Hills 27F, 1-17-1 Toranomon, Minato-ku, Tokyo 105-6427, Japan}

\email{omama@astro.hiroshima-u.ac.jp, omama@ac.jaxa.jp}


\maketitle

\begin{abstract}

Simultaneous X-ray and optical observations of
black hole X-ray binaries have shown that the
light curves contain multiple correlated and
anti-correlated variation components when the
objects are in the hard state.
In the case of the black hole X-ray binary,
GX $339-4$, the cross correlation function
(CCF) of the light curves suggests a positive
correlation with an optical lag of $0.15$ s
and anti-correlations with an optical lag of
$1$ s and X-ray lag of $4$ s.
This indicates the two light curves have some
common signal components with different delays.
In this study, we extracted and reconstructed
those signal components from the data for GX
$339-4$.
The results confirmed that correlation and
anti-correlation with the optical lag are two
common components.
However, we found that the reconstructed light
curve for the anti-correlated component indicates
a positively correlated variation with an X-ray
lag of $\sim +1$ s.
In addition, the CCF for this signal component
shows anti-correlations not only with the
optical lag, but also with the X-ray lag, which
is consistent with the CCF for the data.
Therefore, our results suggest that the 
combination of the two positively correlated 
components, that is, the X-ray preceding signal
with the $0.15$-s optical lag and the optical 
preceding signal with the $1$-s X-ray lag, can 
make the observed CCF without anti-correlated 
signals.
The optical preceding signal may be caused by 
synchrotron emission in a magnetically 
dominated accretion flow or in a jet, while further study
is required to understand the mechanism of the
X-ray time lag.

\end{abstract}

\section{Introduction}

Black hole X-ray binaries (BHXRB) are close
binary systems composed of a black-hole (BH)
and a companion star.
They have two major states, depending on whether
X-ray emissions are soft with a high luminosity
or hard with a low luminosity
\citep{remillard2006x}.
A thermal component dominates the X-ray
spectrum in the soft state.
In this state, it is believed that the system
has an optically thick, geometrically thin disk,
the so-called standard disk, which reaches to
the innermost stable circular orbit
\citep{shakura1973black}.
In contrast, in the hard state, the
X-ray spectrum is dominated by non-thermal
emission.
It is believed that advection-dominated
accretion flow (ADAF) or jets are responsible
for the non-thermal emissions
(\cite{narayan1994advection},
\cite{abramowicz1994thermal},
\cite{fender2004towards}).

Multi-wavelength observations have revealed
various time-lagged correlations between the
light curves for the different wave bands in
BHXRBs.
In the hard state, correlated X-ray and
optical/near-infrared variations are
occasionally observed over short time scales
(0.01-10 s)
(\cite{motch1983simultaneous},
\cite{kanbach2001correlated},
\cite{hynes2009echo},
\cite{gandhi2008rapid},
\cite{Gandhi_2010},
\cite{casella2010fast}
\cite{gandhi2017elevation},
\cite{paice2019black}).
The fast variability has common features in
the cross-correlation function (CCF): 
a positive correlation with an optical time lag
on the orders of 0.1-1.0 s and possible negative
correlations with optical and/or X-ray time
lags longer than the positive correlation lag
(\cite{motch1983simultaneous}, 
\cite{kanbach2001correlated},
\cite{durant2008swift},
\cite{gandhi2008rapid},
\cite{hynes2009echo}, 
\cite{veledina2015discovery},
\cite{paice2019black}).
Such correlations are a key to understanding the
accretion flow and jets in BHXRBs.

It is possible that the reprocessing of X-rays
in the outer disk makes positively correlated
X-ray and optical variations with an optical
lag.
However, the observed time lag on the orders
of 0.1-1.0 s is too short for the reprocessing
scenario
(e.g., \cite{kanbach2001correlated},
\cite{gandhi2008rapid},
\cite{gandhi2017elevation}).
Moreover, the optical delay may be explained
by the travel time of electrons in the jet:
the electrons generate X-rays while they are 
very close to the BH, but they produce optical
emissions after they move downstream in the jet
\citep{gandhi2017elevation}.
The origin of the weak anti-correlations is
unclear (e.g., \cite{veledina2011synchrotron},
\cite{malzac2004jet}).
It is difficult to explain them with a
standard process of energy transfer, such as
the reprocessing of X-rays.

The CCF gives the time lag between the
correlated light curves.
However, examining only the CCF makes it
difficult to investigate the properties of
these light curves.
If the CCF indicates correlation or
anti-correlation, there should exist
corresponding common signal components between
X-ray and optical light curves.
Extracting those signal components will
provide further information on the cause of
the correlated variation.
In this study, we extracted the light
curves for each common signal component with
different time lags based on a time-frequency
analysis.
In our method, we assumed sparsity in the
Fourier basis.
This enables us to effectively select the
Fourier components that are essential for
the common signal components.

GX 339$-$4 is a BHXRB whose mass function is
estimated to be $5.8M_\odot$
(\cite{hynes2003dynamical}, 
\cite{heida2017mass}).
The object exhibits the typical CCF of X-ray
and optical light curves
(\cite{gandhi2008rapid}; hereafter G08,
\cite{Gandhi_2010}).
The CCF has three extrema between $-10$ s and
$+10$ s: a peak at $+0.15$ s, and the troughs
at $+1$ s and $-4$ s, where the positive time
lag is the optical lag with respect to the X-ray
variation.
In this work, we applied our method to the
Night 1 data in G08.

In this paper, section 2 describes the data
and preprocessing.
The details of the signal processing are
introduced in section 3.
Section 4 provides the results.
We discuss the implications of the results
in section 5.
Finally, we summarize our findings in
section 6.

\section{Data and preprocessing}

\begin{figure*}
 \begin{center}
  \includegraphics[width=160mm]
  {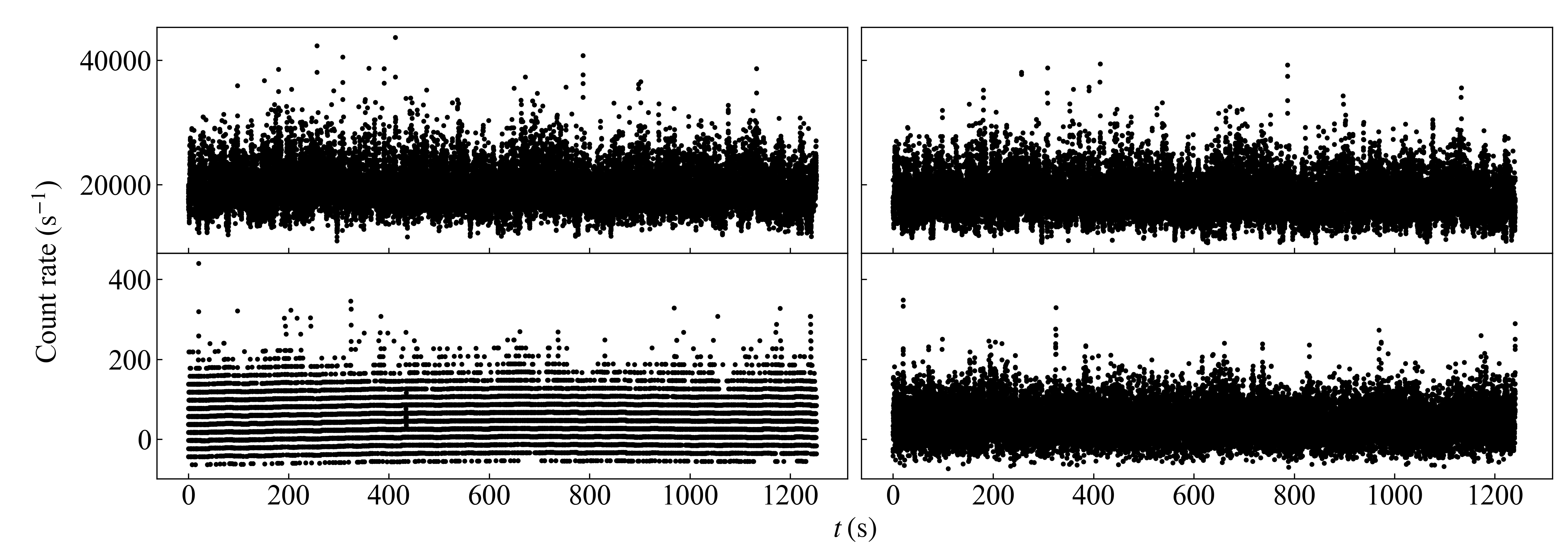}
 \end{center}
 \caption{%
  Left: Observed optical (upper) and X-ray (lower)
  light curves.
  Right: Filtered optical (upper) and X-ray (lower)
  light curves.}%
 \label{fig:lc_obs}
\end{figure*}

We used the same X-ray and optical data
as those for Night 1 in G08.
The observation was on 18 June 2007, when
the object had returned to a low-flux state
after an outburst, and was in the hard state.
The luminosity of 1-100 keV was $5.3 \times
10^{36}\;{\rm erg\, s^{-1}}$, corresponding
to a ratio of 0.007 to the Eddington
luminosity \citep{Gandhi_2010}.
The optical and X-ray data were obtained with
the ULTRACAM attached to the Very Large
Telescope and the Proportional Counter Array
on the Rossi X-ray Timing Explorer satellite,
respectively.
The overlap duration of the observations was
about an hour.
The duration of the Night 1 data was 1200 s.
The time interval of the data was 50 ms;
thus, the number of points in each light
curve is 24000.
The observed optical and X-ray light curves
are shown in the left panels of figure 
\ref{fig:lc_obs}.
In this study, we used the time $t$, which is
the elapsed time from MJD 54269.31472020 in
seconds.
In the X-ray light curve, there are only 
$\lesssim 10$ counts, corresponding to
a count rate of $\lesssim 200$, in most of
the time bins after the background subtraction.
Such a small number of counts means that the
data contains noise due to the Poisson
statistics.

\begin{figure}
 \begin{center}
  \includegraphics[width=80mm]{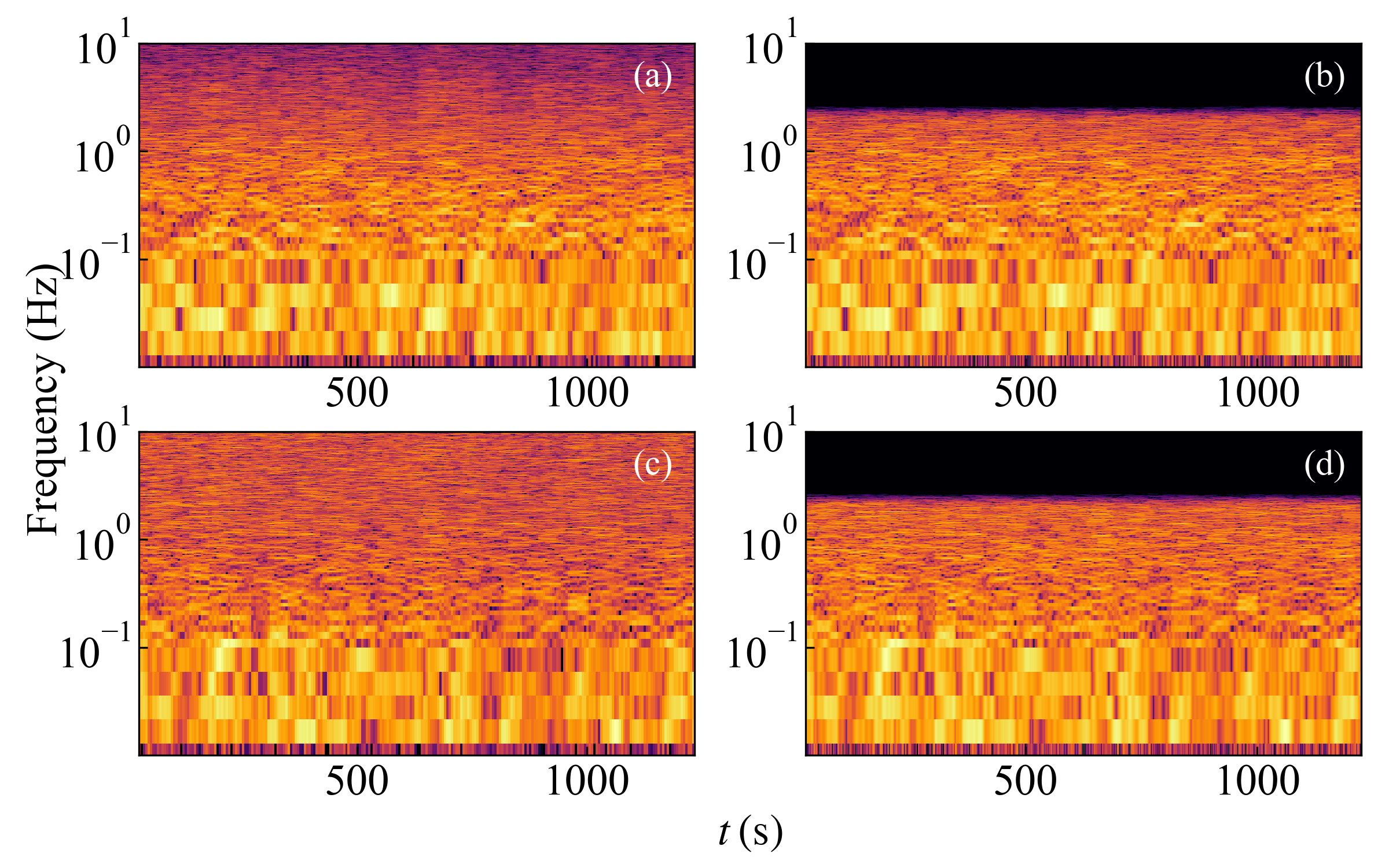}
 \end{center}
 \caption{%
  Spectrograms of the original light curves and
  the filtered light curves.
  The power spectra were calculated for each
  short frame of 50 s.
  We used a Hanning window to reduce spectral
  leakage.
  Panels (a) and (c) are spectrograms of optical and X-ray
  original light curves, respectively.
  Panels (b) and (d) are spectrograms of optical and X-ray
  filtered light curves,  respectively.}%
 \label{fig:spectrogram}
\end{figure}

The spectrograms of the observed optical and X-ray 
data are shown in panels (a) and (c)
of figure \ref{fig:spectrogram}.
We can see significant power in the high
frequency band from $2$ Hz to $10$ Hz in the
X-ray spectrogram, although it is weak in the
optical one.
The Poisson noise is dominant in the
high-frequency band, compared with the
common signal components.
We applied a low pass filter to cut out high
frequency components and noise.
The filtered light curve has less Poisson noise,
and we estimated the Fourier coefficients with
the constrained least square method, assuming
Gaussian noise.
We used a finite impulse response (FIR) filter
with a length of 19, a cutoff frequency of 2 Hz,
and a gain less than $-60$ dB between 3 and
10 Hz.
To cancel the phase shift caused by
the FIR filter, we applied it in the forward
and backward directions.
Because the filter was applied twice, the gain
between 3 and 10 Hz is less than $-120$ dB and
the phase information is preserved.
The same processing was applied to the optical
light curve.
The filtered light curves are shown in the
right panels of figure \ref{fig:lc_obs}.
In panels (b) and (d) of figure
\ref{fig:spectrogram}, we show the optical and X-ray
spectrograms, respectively, of the
filtered light curves.
We can confirm that high frequency noise
is largely reduced in both datasets.
We used the filtered light curves and the
frequencies from 0 to 2 Hz in the following
analysis.

\section{Methods}\label{sec:methods}

\subsection{Common-signal analysis}
\label{ssec:cs}

We explain how we extracted the common signal components with
time lags from two observed light curves. 
This common signal (CS) analysis is based on the Fourier
transform and thus the time lag can be estimated for
each frequency.
We assumes the two light curves are sparse in time-frequency domain.
Here, "sparse" means many coefficients of the time-frequency
components are zero.
\citet{kato2012period} reported the application of a related
approach for the estimation of the power spectrum from a
single light curve.

The inverse Fourier transform gives a light
curve, $h(t)$, from a complex frequency
sequence, $H(f)$:
\begin{equation}
 h(t) = \int_{-\infty}^{\infty}
         H(f)\exp{(-2\pi ift)}df.  
\end{equation}
Here, the mean in $h(t)$ is taken to be zero.
Because we consider real and discrete light
curves, it is rewritten as
\begin{equation}
   h(t_i) = 2\Delta f \sum_{j=1}^M
   [H_{\rm R}(f_j)\cos{(2\pi f_j t_i)} + 
    H_{\rm I}(f_j)\sin{(2\pi f_j t_i)}],
\end{equation}
where each $f_j$ is equally spaced; $\Delta f =
f_{j+1}-f_j$, and $H_{\rm R}(f_j)$ and $H_{\rm
I}(f_j)$ are the real and imaginary parts of
$H(f_j)$, respectively.
Here we assumed $h(t)$ does not include Fourier
components other than $f_1, \cdots f_M$.
This is simply expressed as $\bm{y} =
\bm{A}\bm{x}$ where $\bm{y} = 
[h(t_1), \cdots, h(t_N)]^T$,
$\bm{x} = [H_{\rm R}(f_1), \cdots,
H_{\rm R}(f_M), H_{\rm I}(f_1), \cdots,
H_{\rm I}(f_M)]^T$, and
\begin{equation}
\begin{array}{lll}
 (A)_{i,j}   &=&
  2\Delta f \cos{(2\pi f_j t_i)} \\
 (A)_{i,M+j} &=&
  2\Delta f \sin{(2\pi f_j t_i)}. \\
\end{array}
\end{equation}

We assume that the observed X-ray and optical
light curves have common signal components.
The X-ray and optical light curves,
$\bm{y}_{\rm x}$ and $\bm{y}_{\rm o}$,
respectively, can be expressed with the inverse
Fourier transform as described above,
that is,
\begin{equation}
\begin{array}{lll}
\bm{y}_{\rm x} &=&
 \bm{A}_{\rm x} \bm{x}_{\rm x} \\
\bm{y}_{\rm o} &=&
 \bm{A}_{\rm o} \bm{x}_{\rm o},
\end{array}
\label{eq:DFT_Matrix_xo}
\end{equation}
where $\bm{x}_{\rm x} =  (a_1,\cdots,a_M,b_1,\cdots,b_M)^T$ and
$\bm{x}_{\rm o} = (c_1,\cdots,c_M,d_1,\cdots,d_M)^T$ are the
Fourier components, and $\bm{A}_{\rm x}$ and $\bm{A}_{\rm o}$
are the matrices of the inverse Fourier transform.
Here, $\{H_R(f_j), H_I(f_j)\}$ are denoted by $\{a_j, b_j\}$ and
$\{c_j, d_j\}$ for the X-ray and optical light curves,
respectively.
We also note that $\bm{y}_{\rm x}$ and $\bm{y}_{\rm o}$ are
standardized to a zero mean and unit variance.
The matrices in equation (\ref{eq:DFT_Matrix_xo})
can be expressed by a single equation:
$\bm{y}_{\rm xo} = \bm{A}_{\rm xo}
\bm{x}_{\rm xo}$,
where $ \bm{y}_{\rm xo} =
(\bm{y}_{\rm x}^T,\bm{y}_{\rm o}^T)^T$,
$\bm{x}_{\rm xo} = (\bm{x}_{\rm x}^T,
\bm{x}_{\rm o}^T)^T$, and
\begin{equation}
\begin{array}{ll}
\bm{A}_{\rm xo} = \left(
                   \begin{array}{cc}
                    \bm{A}_{\rm x} & O \\
                    O & \bm{A}_{\rm o} \\ 
                   \end{array}
                  \right).
\end{array}
\label{eq:DFT_Matrix}
\end{equation}
In CS analysis, we estimate the Fourier coefficients,
$\bm{x}_{\rm xo}$ as follows:
\begin{equation}
   \hat{\bm{x}}_{\rm xo}
    = \arg \min_{\bm{x}_{\rm xo}}
    \left[||\bm{y}_{\rm xo}
    - \bm{A}_{\rm xo}\bm{x}_{\rm xo}||_2^2 
     + \lambda \sum_j
    \sqrt{a_j^2+b_j^2+c_j^2+d_j^2} \right], \\
   \label{eq:common-signal}
\end{equation}
\begin{equation}
   \hat{\bm{x}}_{\rm xo}
    = {\rm CS}(\bm{y}_{\rm xo}), \\
\end{equation}
where $\lambda$ is the penalty coefficient
that controls the sparsity of
$\hat{\bm{x}}_{\rm xo}$.
The form of equation (\ref{eq:common-signal})
is known as the group lasso
\citep{yuan2006model}.
By grouping the X-ray and optical
coefficients, it is expected that the
coefficients become zero for the Fourier
components that have high power only in
either the X-ray or optical light curve, and
that the relevant components of the common
signals will be extracted.
Note that even if the X-ray and optical signal
components with the same frequency are extracted,
their phases can be different because the grouped
signals include sine and cosine simultaneously.
Therefore, we can analyze their time lags.

In this study, we chose $\lambda$ using the
$k$-fold cross-validation method.
The model was evaluated with the mean squared
error yielded by the prediction from the
training model on the validation dataset.
We determined $\lambda$ at which the mean
squared error was minimized.
We set $k=10$.

Using $a_j$ and $b_j$ for the X-ray data, and
$c_j$ and $d_j$ for the optical data, we can
calculate the amplitudes, $Q_{{\rm x},j}$ and
$Q_{{\rm o},j}$, and the phases $\phi_{{\rm
x},j}$ and $\phi_{{\rm o},j}$, of a frequency,
$f_j$ as follows:
\begin{equation}
   \begin{array}{l}
   Q_{{\rm x},j} = \sqrt{a_j^2+b_j^2}, \\
   Q_{{\rm o},j} = \sqrt{c_j^2+d_j^2},
   \end{array}
\end{equation}
\begin{equation}
   \begin{array}{l}
   \phi_{{\rm x},j} =
    2\pi f_j\Delta t_{{\rm x},j} =
    \arctan{(b_j/a_j)}, \\
   \phi_{{\rm o},j} =
    2\pi f_j\Delta t_{{\rm o},j} =
    \arctan{(d_j/c_j)}.
   \end{array}
\end{equation}
We obtain the time lag of the optical variation
with respect to the X-ray one, $\tau_j$, as
follows:
\begin{equation}
 \tau_j = 
  \left\{
  \begin{array}{ll}
  \Delta t_{{\rm x},j} -
  \Delta t_{{\rm o},j} \
   (-T_j/2 \leq \Delta t_{{\rm x},j} -
   \Delta t_{{\rm o},j} < T_j/2),\\
  T_j/2 - (\Delta t_{{\rm x},j} -
  \Delta t_{{\rm o},j}) \
   ({\rm otherwise}),\\
 \end{array}
 \right.
\label{eq:tau}
\end{equation}
where $T_j$ is the period of the $j$-th
Fourier component, $T_j=1/f_j$.
Note that $\tau_j$ is defined as being between
$-T_j/2$ and $T_j/2$.

\subsection{Identification of the Fourier components for time lags}
\label{ssec:identification}

\begin{figure*}
\begin{minipage}{0.5\hsize}
 \begin{center}
  \includegraphics[width=85mm]{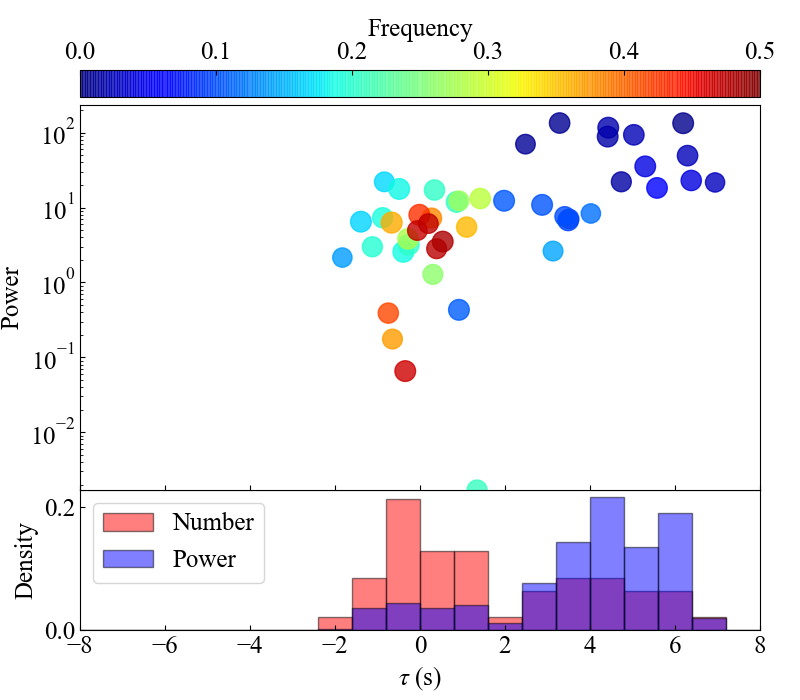}
 \end{center}
\end{minipage}
\begin{minipage}{0.5\hsize}
 \begin{center}
  \includegraphics[width=85mm]{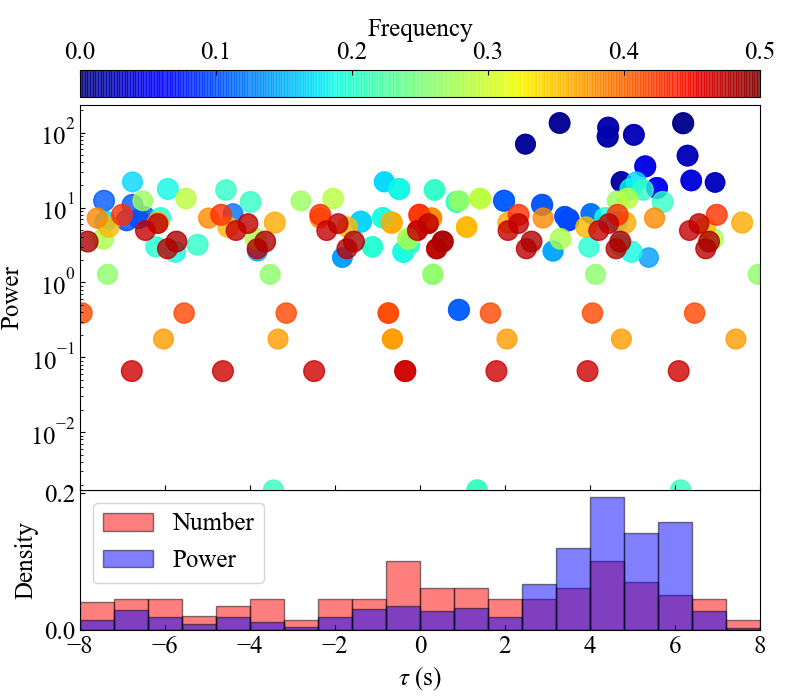}
 \end{center}
\end{minipage}
\caption{%
 Examples of the results of CS analysis. 
 The data are the simulated set with a true
 time lag of 5 s.
 Upper panel: color-bubble plot of the
 Fourier components.
 The horizontal axis is the lag, and the
 vertical axis is the amplitude.
 The color and size of the bubbles represent
 the frequency and the amplitude ratio, $r$
 (for details, see the text).
 Lower: histogram of the Fourier components
 (red), and the sum of the amplitudes for each
 bin (blue).
 The left and right panels show the results
 without and with periodic counterparts of
 the components, respectively.
 }%
\label{fig:cbp_art}
\end{figure*}

The results of CS analysis are visualized by a
color-bubble plot and its histogram.
An example of the plots obtained with the
simulated data with $\tau=5$ s is shown
in the left panel of figure \ref{fig:cbp_art}.
The simulated X-ray data were a sample drawn from the
algorithm proposed in \citet{timmer1995generating} where 
the power spectrum is expressed with a power law,
and we set $P(\nu)\propto \nu^{-2}$.
The optical light curve is identical to that
for the X-ray curve, but with a time lag of 5 s.
The sampling interval was 1 s.
Each light curve has 100 data points.
Both light curves were standardized to a zero
mean and unit variance.
In the color-bubble plot, a bubble represents
the features of the Fourier component.
The horizontal axis is $\tau$, the vertical
axis is $Q = \sqrt{Q_{{\rm x}}^2 +
Q_{{\rm o}}^2}$, the color indicates the
frequency, and the size indicates the amplitude
ratio, that is,
\begin{equation}
r = \left\{
     \begin{array}{ll}
       Q_{{\rm x}} / Q_{{\rm o}} &
       (Q_{{\rm x}} \leq Q_{{\rm o}}) \\
       Q_{{\rm o}} / Q_{{\rm x}} &
       (Q_{{\rm x}} > Q_{{\rm o}}).
     \end{array}
    \right.
\label{eq:ratio}
\end{equation}
A small $r$ means that the component has a
high power only in one of the light curves
and implies that it is not a common signal
component.
We only show components which are $r>0.8$, thus the difference
of the size is small in figure \ref{fig:cbp_art}.
In the left panel of figure \ref{fig:cbp_art},
we can see a concentration of components
at $\tau \sim 5$ s as expected.

In general, the true time lag of the $j$-th
Fourier component is indistinguishable from its
periodic counterparts, $\tau_j^k$:
\begin{equation}
 \tau_j^k = \tau_j + kT_j\ 
 (k=\cdots,-1,0,+1,\cdots).
\label{eq:tau_periodical}
\end{equation}
In equation (\ref{eq:tau}), we define the range
of $\tau_j$ as  $-T_j/2 \leq \tau_j < T_j/2$.
However, it is possible that the true time lag
is outside this range.
In the left panel of figure \ref{fig:cbp_art},
we can see some concentrations of the Fourier
components that appear due to the ambiguity in
the time lag and our definition of the $\tau$
range.
In addition to the concentration around the
true time lag of $\tau=5$ s, we can also see
another concentration at $\tau \sim 0$ s, which
apparently implies a common signal component
without a significant time lag.
The amplitudes of the components at $\tau
\sim 0$ s are smaller than those around the
true time lag.
The concentration at $\tau \sim 0$ s is made
by the components of high frequencies.
The right panel of figure \ref{fig:cbp_art}
shows the bubble plot including the
periodic counterparts with each $T_j$.
We can see that most of the high-frequency
components have counterparts around $\tau=5$ s.
Thus, we should select the time lag of each
Fourier component from one of the periodic
counterparts, and should extract Fourier
components to reconstruct the light curve.
When we focus on a concentration close to
$\tau \sim 0$ s, it blends with the
spurious concentration made by the high
frequency components.
In this case, we can check the significance of
the concentration by a statistical test, as
described in Appendix \ref{ap:signiftest}.
In the left panel of figure \ref{fig:cbp_art},
there is another weak concentration at
$\tau \sim -3$ s, which is nearly opposite in
terms of sign to the true time lag.
As well as the high-frequency components
around $\tau=0$ s, most of the components
around $\tau=-3$ s also have counterparts around
$\tau=5$ s, as shown in the right panel.
These are the components having $T_j/2\sim 5$ s,
in which the time lag of $\tau=+5$ s is almost
identical to that of $\tau=-5$ s.
Hence, we should note that a hint of time lag
would appear around the sign opposite to the
true time lag.

We calculated $\tau_j$ assuming a positive
correlation between the X-ray and optical light
curves, whereas anti-correlated variations were
also reported in the G08 data.
The time lag in anti-correlation,
$\tau^{\rm anti}_j$, can be obtained from
$\tau_j$ and the period, $T_j$.
When we consider a single frequency wave, for
example, a positive correlation with a positive
$\tau_j$, can also be interpreted as an
anti-correlation with a negative
$\tau^{\rm anti}_j$.
Thus, we obtain $\tau^{\rm anti}_j$ by shifting
$\tau_j$ by half of $T_j$,
\begin{equation}
   \tau^{\rm anti}_j = 
    \left\{
     \begin{array}{ll}
     \tau_j - \frac{T_j}{2}\ 
     &(\tau_j \geq 0) \\
     \tau_j + \frac{T_j}{2}\ 
     &(\tau_j < 0).
  \end{array}
  \right.
  \label{eq:AntiCorr}
\end{equation}

\subsection{Short-time common signal analysis (STCS)}
\label{ssec:stcs}

We performed the short time common-signal (STCS)
analysis, in which we divided the
light curves into short segments and conducted
CS analysis on them.
STCS analysis is essential for reconstructing the light
curves when the common signals are stationary
for short-time intervals but non-stationary for
longer durations.
This is similar to the time-frequency analysis
for acoustic signal processing.

The $m$-th short segment, $y_m(t)$, is given
from the whole light curve $y(t)$, as follows:
\begin{equation}
y_m(t-mS) = w_s(t-mS)y(t),
\end{equation}
where $S$ is the frame shift in seconds and
$w_s(t)$ is the window function.
Each segmented light curve has a time range
of $mS \leq t < (mS+\ell)$, where $\ell$ is
the frame length in seconds.
In the case of the observed data, the $m$-th
segment can be expressed as $\bm{y}_m =
(y_m(t_1-mS), \cdots, y_m(t_N-mS))$, where
$N$ is the number of samples in each frame.
We estimate the Fourier coefficients for the
$m$-th frame, $\bm{\hat{x}_m}$, with CS analysis, that
is,
\begin{equation}
   \bm{\hat{x}_m} =
    {\rm CS}(\bm{y_m}).
\end{equation}
We used $\ell=50$ s, $S=1$ s, and $N=1000 (\ell \times 20
\mathrm{Hz})$ for the data of GX 339$-$4 in this study.
The number of frames was 1200
($m=1, 2, \cdots, 1200$).
We used the Hanning window function as
$w_s(t)$.
In short time Fourier analysis, window functions are employed in
order to prevent power leakage in the frequency domain.
The Hanning window is one of the popular window functions.
We set the window size same as the length of the frame, that is,
50 s.

\subsection{Reconstruction of light curves from STCS analysis}

We reconstructed the common light curves
with a given time lag based on the inverse
Fourier transform.
Each reconstructed segment is expressed as
\begin{eqnarray}
   \tilde{y}_m(t)
   &=& \frac{1}{N} \sum_{k \in K}
       \hat{x}_{m,k} \exp{(2\pi if_kt}) \\
   &=& \frac{1}{N} \sum_{k \in K}
       [a_k\cos{(2\pi f_kt)} +
        b_k\sin{(2\pi f_kt)}],
\end{eqnarray}
where $K$ represents a subset of the Fourier
components selected for the given time lag.
The light curve is reconstructed by superposing
all reconstructed segments multiplied by the
weighting function,
$w_a(t)$.
\begin{equation}
   \hat{y}(t) =
    \sum_m w_a[t-(mS+\ell/2)]\tilde{y}_m(t),
\end{equation}
where $t-(mS+\ell/2)$ is the middle time of
each segment.
We used a triangle-shaped window as $w_a(t)$:
\begin{equation}
w_a(t') = \left\{
 \begin{array}{ll}
  \frac{4S}{d^2}t' + \frac{2S}{d}  \ 
   &(-d/2 \leq t' \leq 0)\\
  -\frac{4S}{d^2}t' + \frac{2S}{d} \ 
   &(0 < t' \leq d/2) \\
  0 \ &({\rm otherwise}), \\
 \end{array}
 \right.
\label{eq:w_a}
\end{equation}
where $d$ is the length of the extracted range.
Because it is known that the signals are not reliably
reconstructed around the edges of each windowed frame,
only the center part of $\tilde{y}_m(t)$ was used for $\hat{y}(t)$.
We set $d=2$ s.

\section{Results}
\label{sec:results}

\begin{figure*}
\begin{minipage}{0.5\hsize}
 \begin{center}
  \includegraphics[width=85mm]{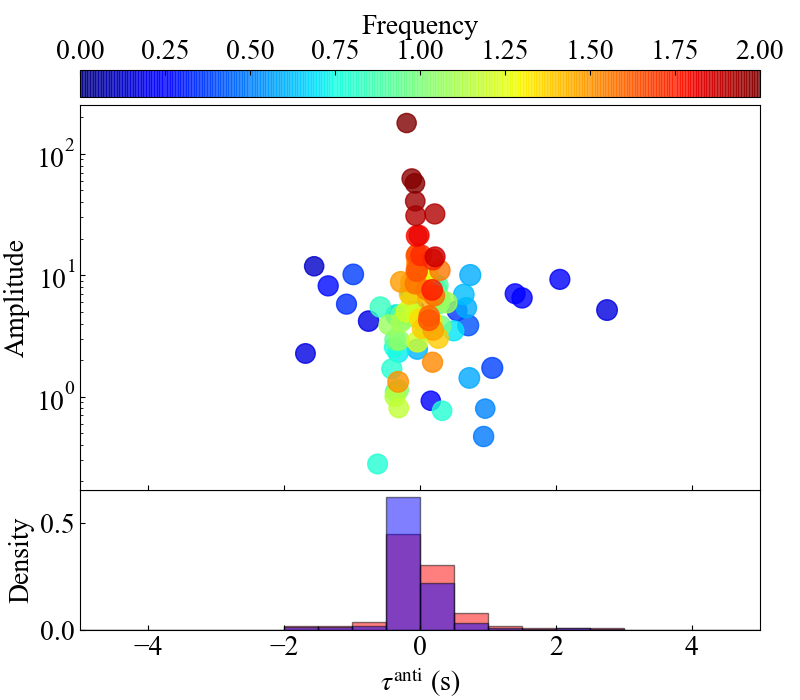}
 \end{center}
\end{minipage}
\begin{minipage}{0.5\hsize}
 \begin{center}
  \includegraphics[width=85mm]{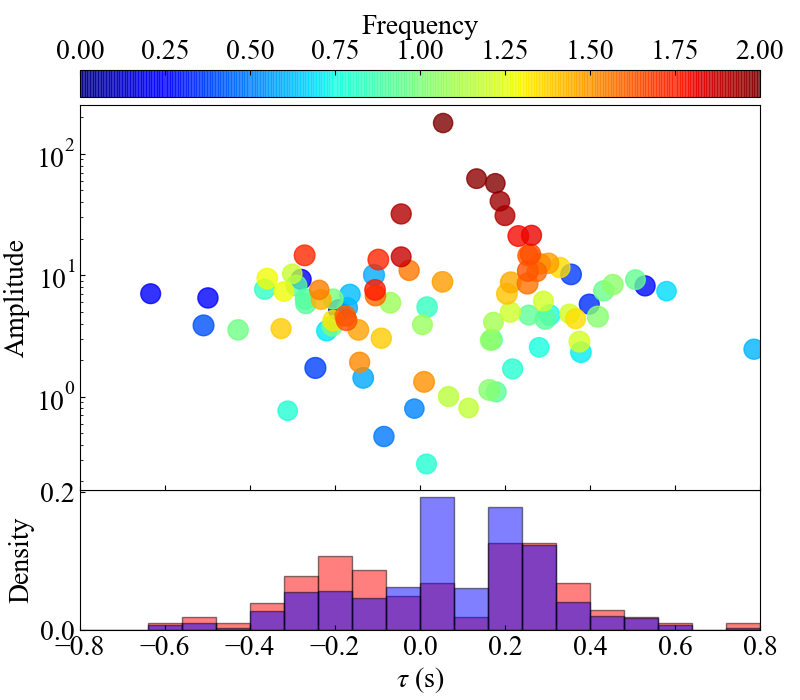}
 \end{center}
\end{minipage}
\caption{%
 Bubble plots for the frame $t=615$-$665$.
 The horizontal axes of the left and right
 panels are $\tau^{\rm anti}$ and $\tau$,
 respectively.}%
\label{fig:omplot}
\end{figure*}

We performed an STCS analysis for the filtered
data for GX $339-4$.
We show an example of the bubble plots obtained
with each frame in figure \ref{fig:omplot}.
These are the results for the frame between
$t=615$ s and $665$ s.
We note that the horizontal axes of the left
and right panels are $\tau^{\rm anti}$ and
$\tau$, respectively.
The figure only contains high-value $r$
components, that is, $r>0.8$.
The dependence of the results on the threshold of $r$ is
discussed in Appendix 2.
G08 reported an anti-correlation with
$\tau^{\rm anti}=+1$ s and $-4$ s.
In the left panel, the high-frequency components
are prominently concentrated at $\tau^{\rm anti}
\sim 0\;{\rm s}$, as described in section
\ref{ssec:cs}.
In addition, weak excesses can be seen in
$\tau^{\rm anti}$ between $-2.0$ and $-0.5$ s and
between $+0.5$ and $+3.0$ s.
As described in section \ref{ssec:cs}, they
may correspond to a single time lag.
Regarding a sub-second scale, G08 reported 
$\tau=0.15$ s.
In the right panel, we can see the peaks in the
histogram at $\tau \sim 0.25$ s and $-0.2$ s.
They may correspond to a single time lag of
$\tau=0.15$ s in G08.

\begin{figure*}
 \begin{center}
  \includegraphics[width=160mm]{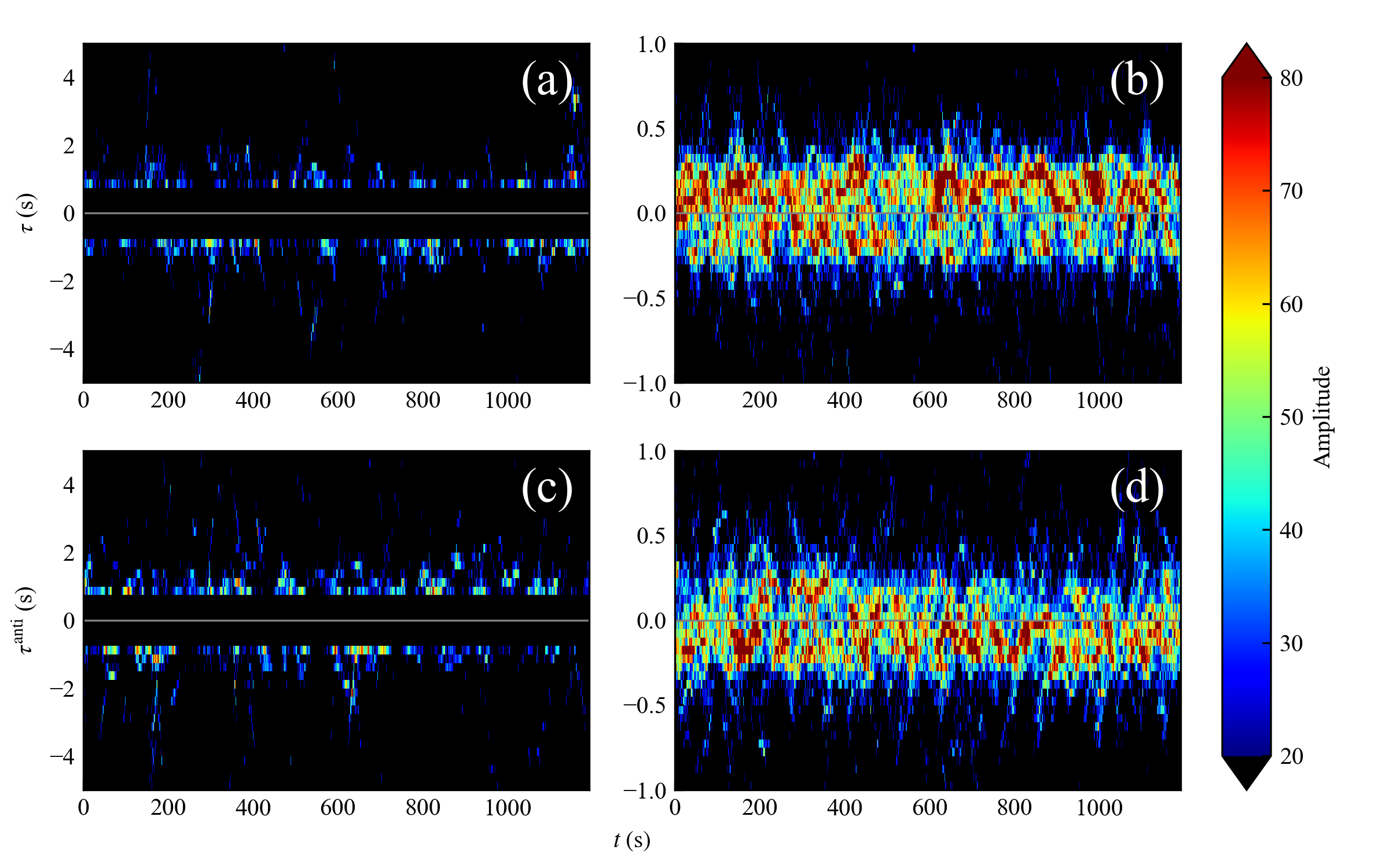}
 \end{center}
 \caption{%
  Time lag maps obtained with the STCS analysis.
  The vertical axes indicate $\tau$ and
  $\tau^{\rm anti}$ in panels (a), (b), and
  (c), (d), respectively.
  The color represents the total amplitude in
  each bin.
  We masked the band between $-0.8$ and $0.8$
  s in panel (a) and (c) to emphasize the
  structure in the region of long time lags.}%
 \label{fig:lagmap}
\end{figure*}

Figure \ref{fig:lagmap} shows the temporal
variation of the time lag obtained from the
STCS analysis.
The horizontal and vertical axes are $t$ and
$\tau$ or $\tau^{\rm anti}$, respectively.
The color indicates the total amplitude of the
Fourier components in each bin, which is
equivalent to the blue boxes in the bubble plot.
We call such a plot a time lag map in this
paper.
Panels (a) and (c) show the time lag map in
$\tau$ and $\tau^{\rm anti}$, respectively.
We masked the time lag between $-0.8$ and
$0.8$ s to emphasize the second-scale time
lags.
No clear concentration of the components can
be seen in the maps, whereas some intermittent
concentrations are evident at $\tau^{\rm anti}
\sim 0.5$-$2.5$ s in panel (c).
This concentration range corresponds to the
anti-correlation with the optical lag of $+1$ s
reported in G08.
We note that there are no hints of component
concentrations at $\tau^{\rm anti}\sim-4$ s,
which was reported in G08.
Panels (b) and (d) are the same as panels (a)
and (c), but for sub-second time lags.
We can see more components in the region
between $\tau=0.1$ s and $0.3$ s than those
between $\tau=-0.1$ and $-0.3$ s in panel (b).
This asymmetry of the distribution corresponds
to the time lag of $\tau=0.15$ s reported
in G08.
In contrast, in panel (d), there is an
opposite asymmetry of the distribution.
They may be the same common signal component
if the time scale of the signal is almost double
that of the time lag.

\begin{figure*}
\begin{minipage}{0.5\hsize}
 \begin{center}
  \includegraphics[width=80mm]{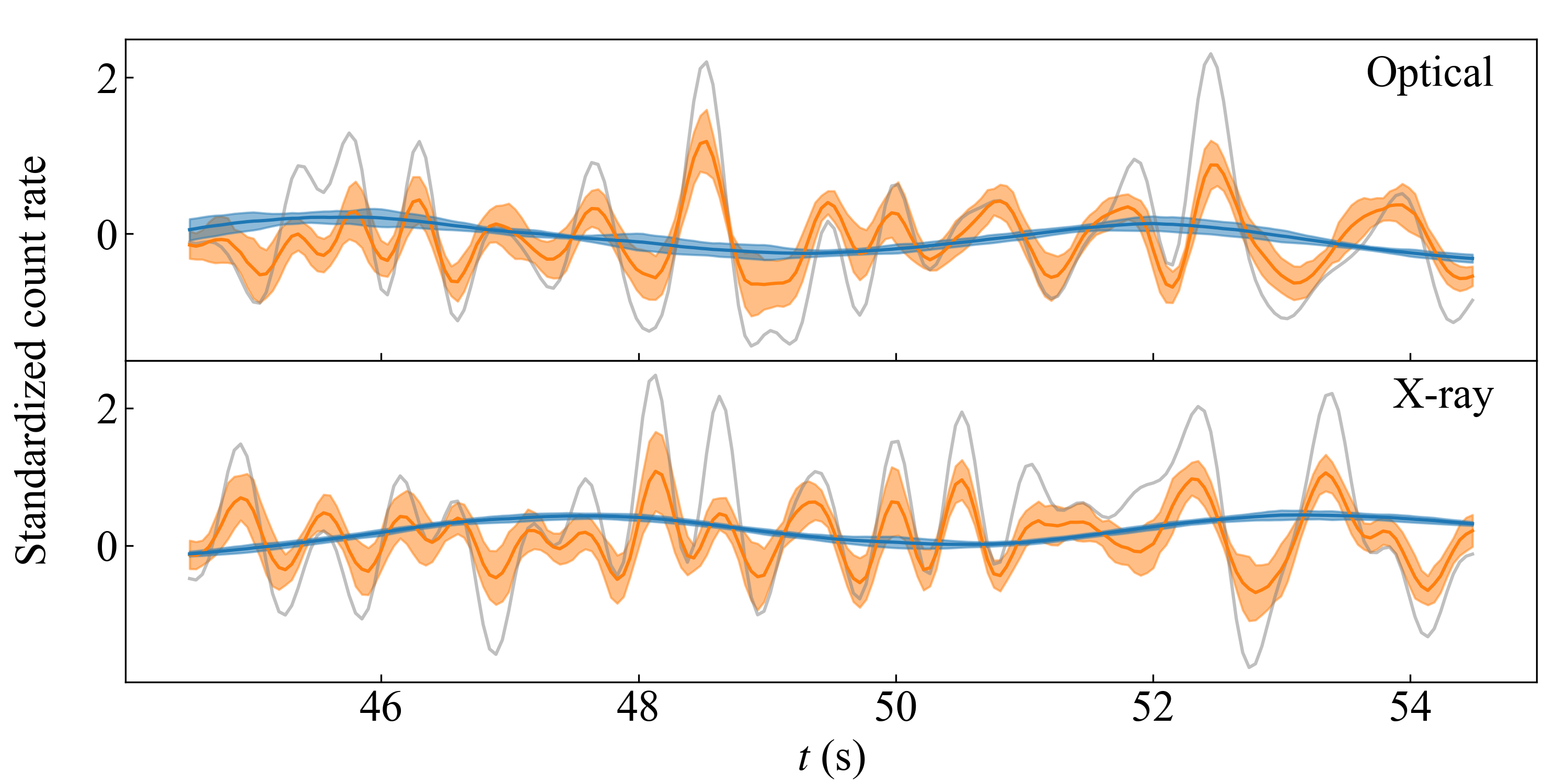}
 \end{center}
\end{minipage}
\begin{minipage}{0.5\hsize}
 \begin{center}
  \includegraphics[width=80mm]{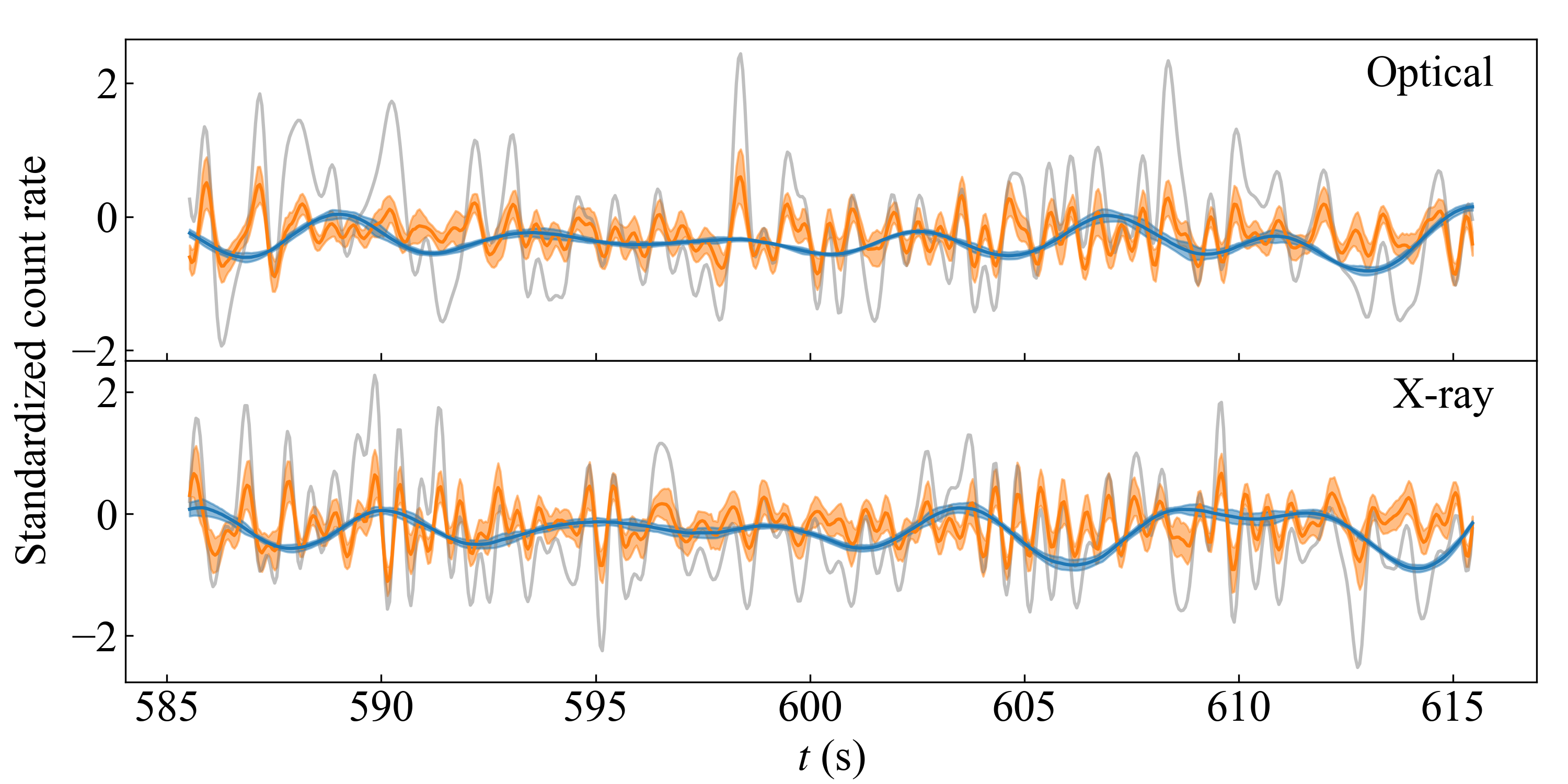}
 \end{center}
\end{minipage}
\begin{minipage}{0.5\hsize}
 \begin{center}
  \includegraphics[width=80mm]{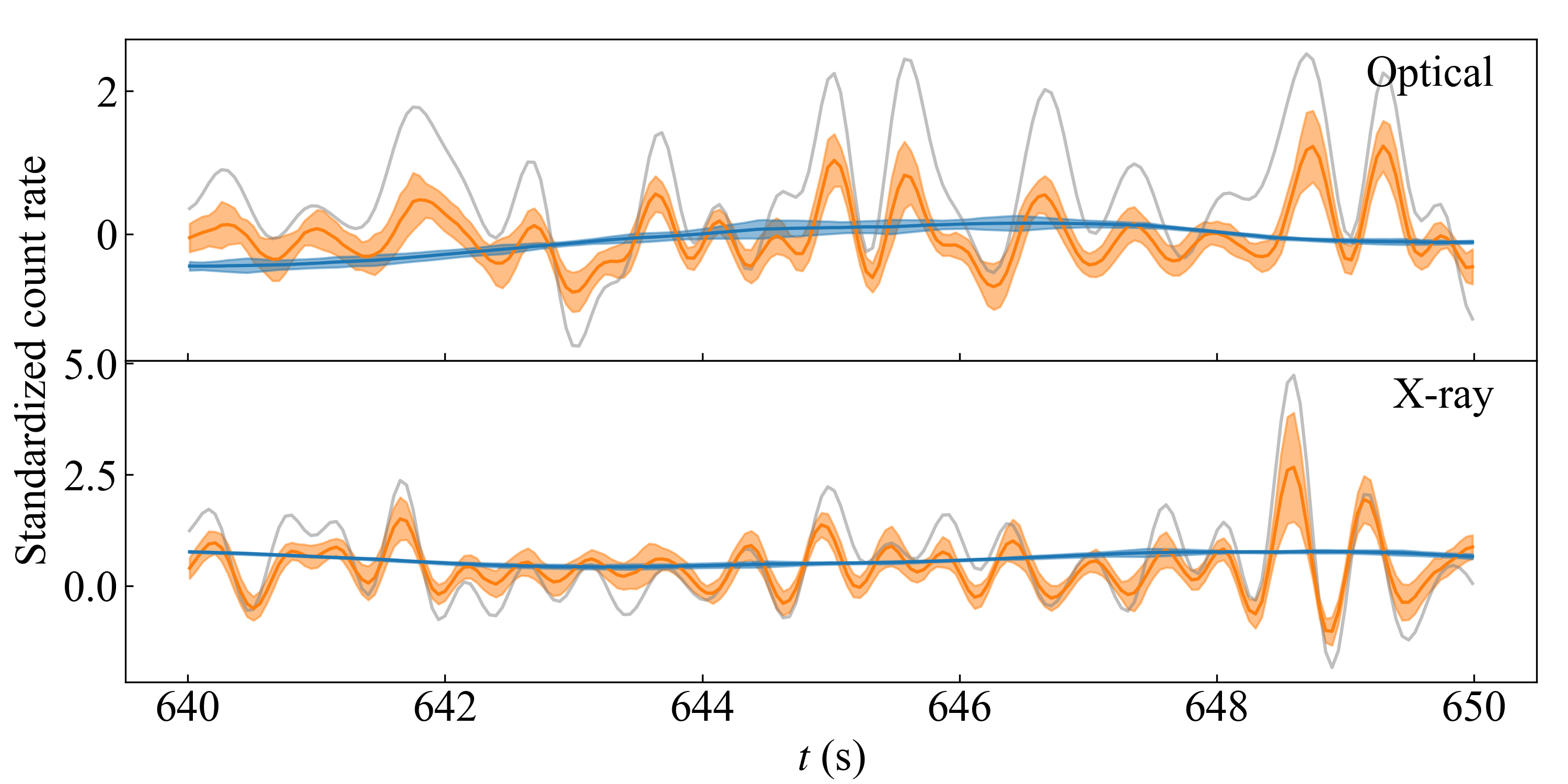}
 \end{center}
\end{minipage}
 \begin{minipage}{0.5\hsize}
 \begin{center}
  \includegraphics[width=80mm]{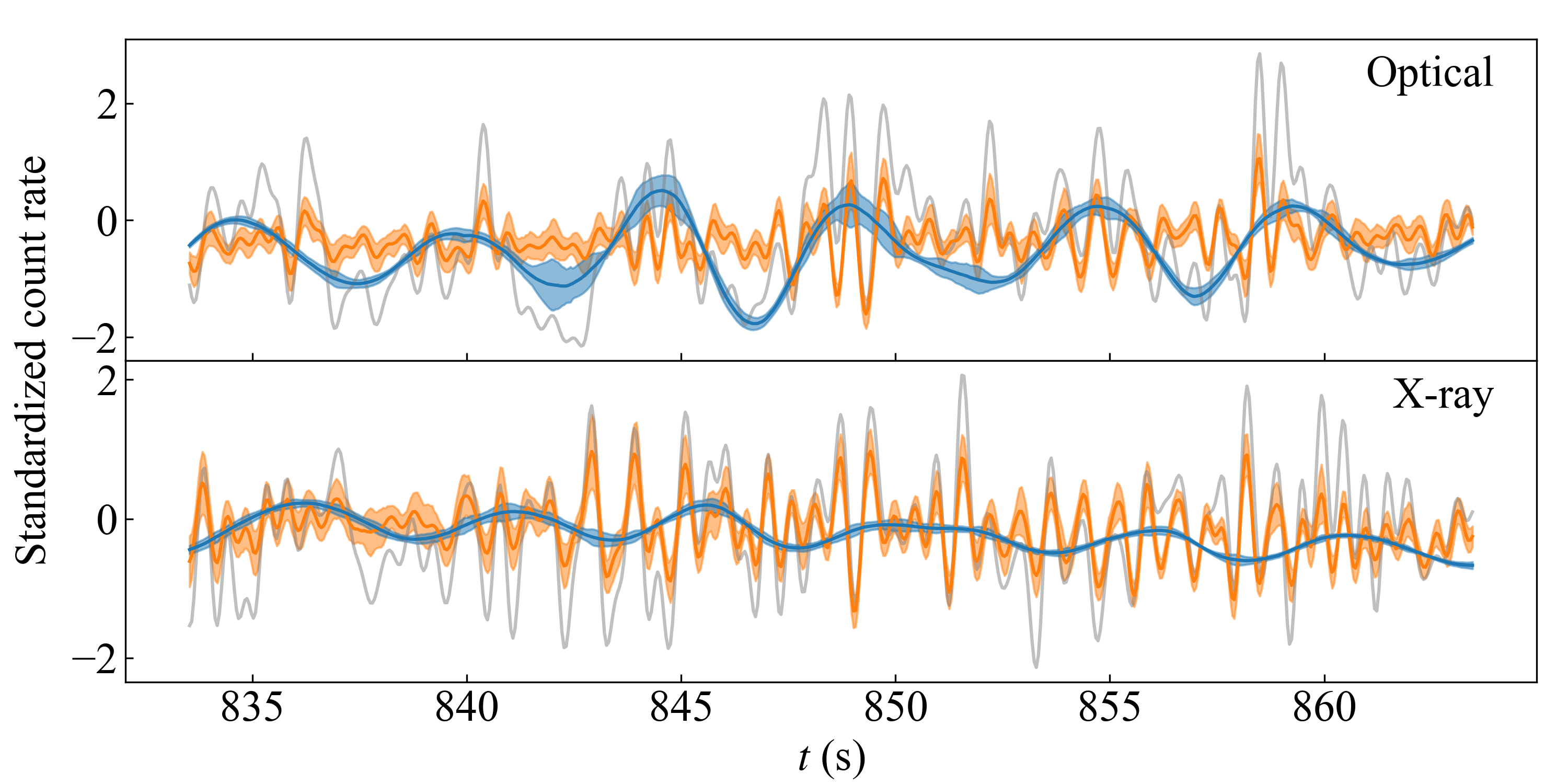}
 \end{center}
\end{minipage}
 \begin{minipage}{0.5\hsize}
 \begin{center}
  \includegraphics[width=80mm]{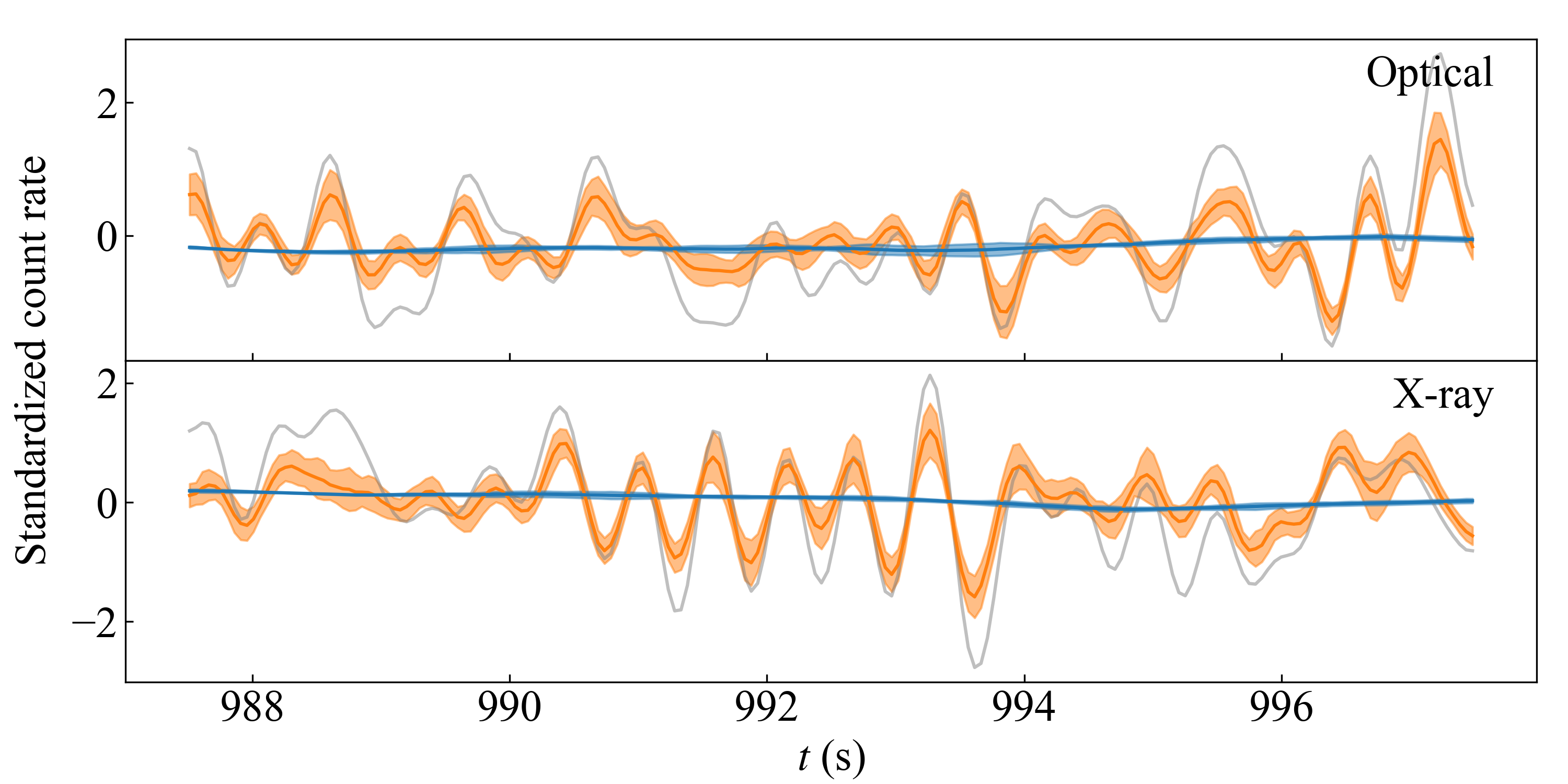}
 \end{center}
\end{minipage}
 \begin{minipage}{0.5\hsize}
 \begin{center}
  \includegraphics[width=80mm]{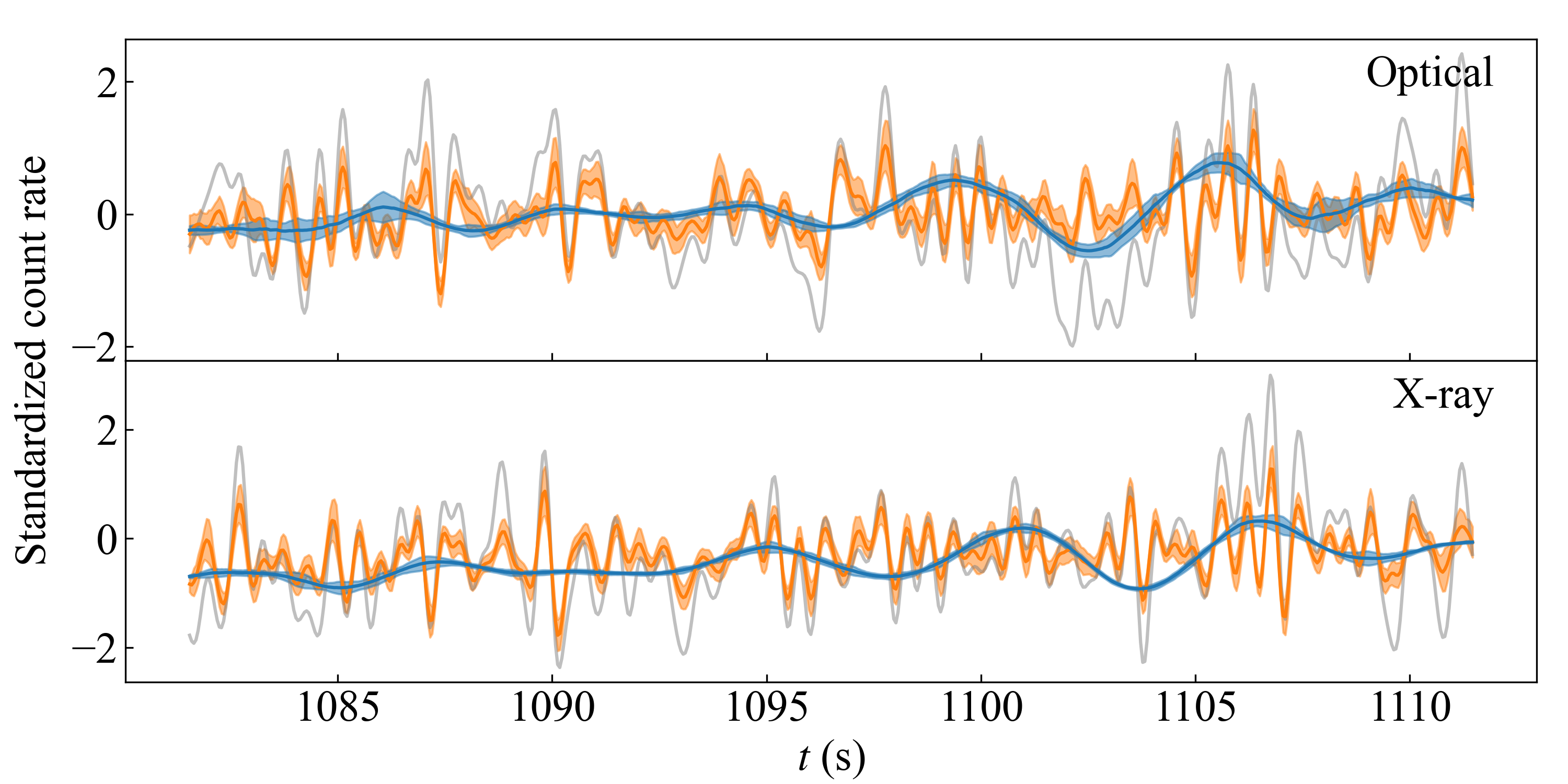}
 \end{center}
\end{minipage}
\caption{%
 Reconstructed light curves.
 The orange, blue, and grey curves are 
 the reconstructed light curve of $\tau=+0.1$ s
 and $\tau^{\rm anti}=+1$ s, and
 filtered light curves, respectively.
 The shaded area represents the 90 \% confidence interval which
 is evaluated via resapmling.
 The upper and bottom parts of each panel are
 the optical and the X-ray light curves,
 respectively.
 The left panel has a short range of 10 s,
 in which the variation of $\tau=+0.1$ is
 clearly seen.
 The right panel has a long range of 30 s, in
 which the variation of $\tau^{\rm anti}=+1$ s
 is clearly seen.
 The count rate of OPS is magnified by 1.5 times to see
 variations clearly.}%
\label{fig:lc_rec}
\end{figure*}

We extracted the Fourier components for the
two common signal components, and reconstructed
their light curves.
In order to test the reliability of the reconstructions, we
evaluated the distribution of each point of the light curve
via resampling.
We randomly selected $K$ samples from $N$ observed data points
and repeated it for $M$ times.
The resulting fluctuation of the reconstructed point gives the
estimation of the distribution.
The result is shown in figure \ref{fig:lc_rec} where $M=100$ and
$K/N = 0.5$.
First, we reconstructed the light curve from
the Fourier components around $\tau=+0.1$ s,
which were extracted by using the significance
test described in Appendix \ref{ap:signiftest}.
We used $\tau$ in a range between $-0.8$ and
$+0.8$ s for the test.
Parts of a reconstructed light curve are
shown in figure \ref{fig:lc_rec} as orange
curves.
We can see that the optical variation is
delayed by $\sim +0.1$ s compared with the
X-ray variation, as expected.

Second, we extracted all the components
between $\tau^{\rm anti}=+0.5$ and
$+2.5$ s for the reconstruction of the light
curves of $\tau^{\rm anti}\sim+1$ s.
We also extracted the periodic counterparts
of the other components that lie in this
$\tau^{\rm anti}$ region.
Parts of the reconstructed light curve are
shown by the blue curves in figure
\ref{fig:lc_rec}.
We found that, although we extracted the
components with a positive $\tau^{\rm anti}$,
the reconstructed light curve can
be interpreted as a positively correlated
variation with an X-ray delay, that is,
$\tau\sim-1$ s.

Thus, our analysis suggests that the
observed light curves include two common
signal components having $\tau \sim +0.1$ s
and $\tau \sim -1$ s.
The signal component of $\tau=+0.1$ s means
that the X-ray variation precedes the optical
variation, thus, we call it the X-ray preceding
signal (XPS).
Likewise, the component of $\tau=-1$ s is
called optical preceding signal (OPS).

\begin{figure}
 \begin{center}
  \includegraphics[width=80mm]{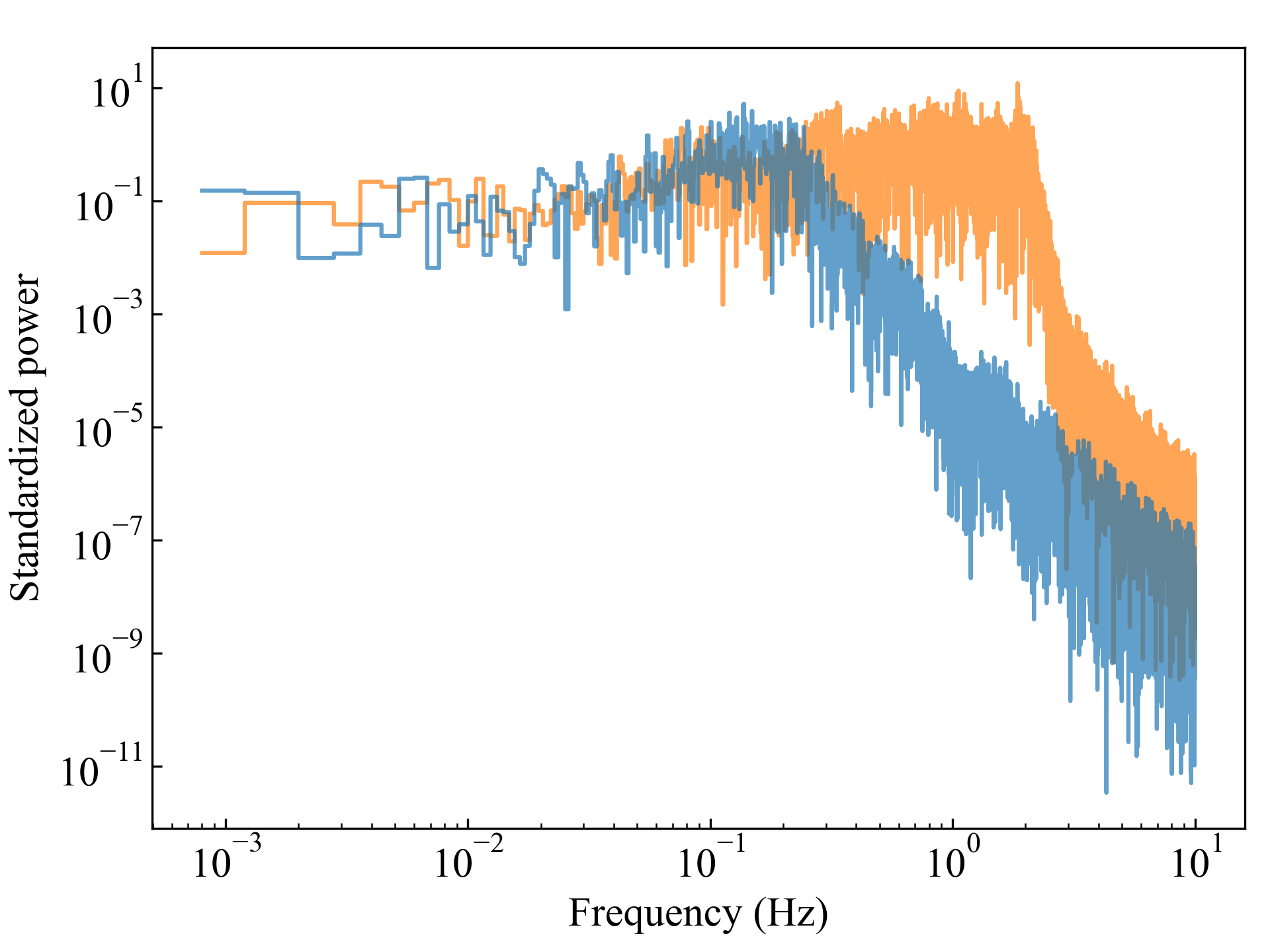}
 \end{center}
 \caption{%
  Power spectrum densities of the XPS and OPS in the optical band,
  indicated by orange and blue curves,
  respectively.}%
 \label{fig:psd}
\end{figure}

The power spectrum densities (PSDs) for the
XPS and OPS in the optical band are shown in
figure \ref{fig:psd} with orange and blue
curves, respectively.
The peak frequencies are $\sim 2$ Hz and
$\sim 0.2$ Hz in PSDs of the XPS and OPS,
respectively. 
We note that the frequency of the XPS only
gives a lower limit of the characteristic
frequency because the FIR filter has a cutoff
frequency of 2 Hz.
Hence, the characteristic time scale of the
XPS and OPS are estimated to be $<0.5$ s and
$\sim 5$ s, respectively.

\begin{figure*}
 \begin{center}
  \includegraphics[width=160mm]{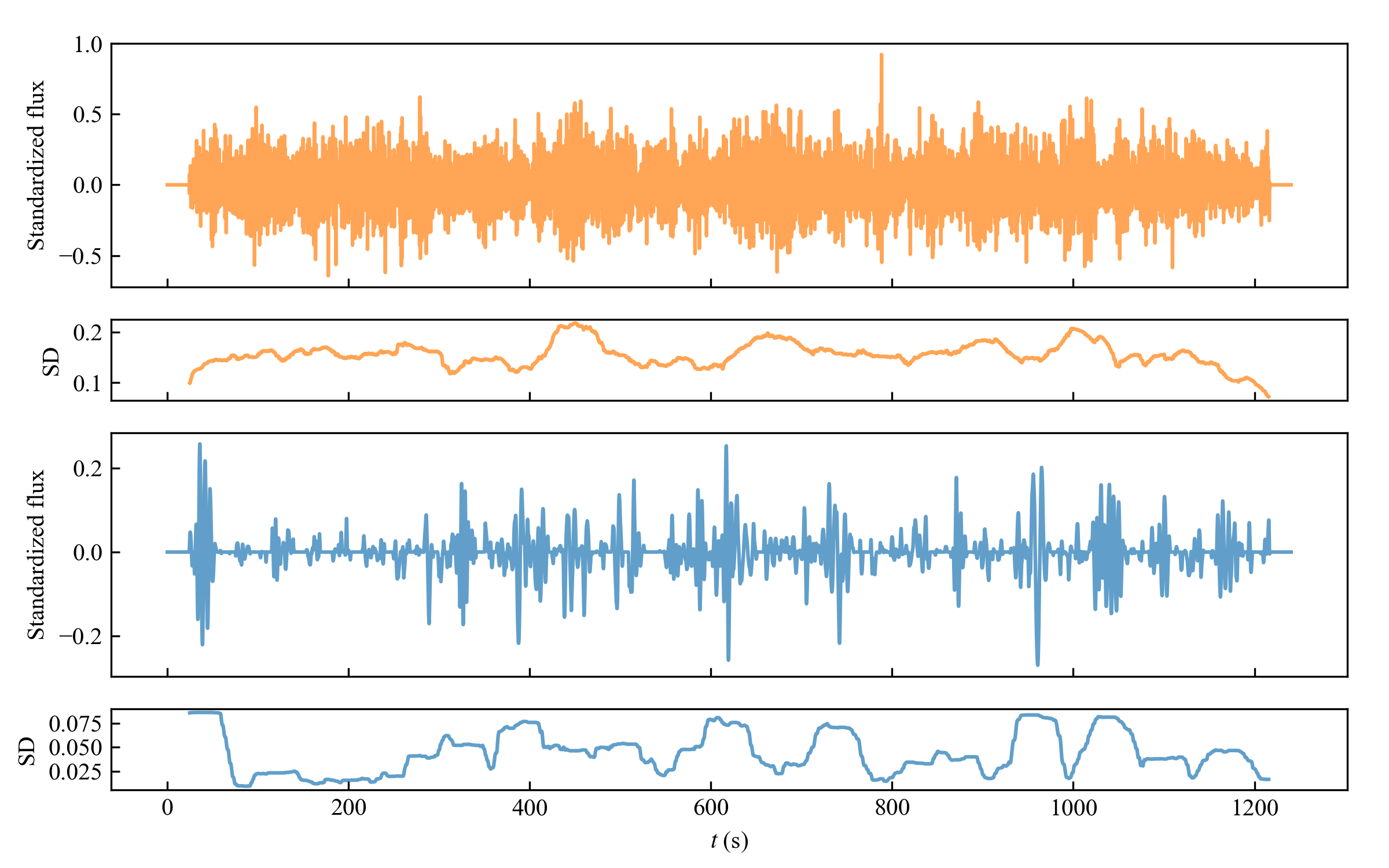}
 \end{center}
 \caption{%
  Reconstructed X-ray light curves for the XPS and OPS, and the
  time variation in the standard deviation (SD) of each frame.
  The orange curves show the XPS light curve and its SD.
  The blue curves show the OPS light curve and its SD.}%
 \label{fig:lcNstd}
\end{figure*}

We show the whole X-ray light curves for the
XPS and OPS in figure \ref{fig:lcNstd}.
The figure includes the time variation of
the standard deviation (SD) of the light
curve in each 50-s frame.
We can see that the active periods for the
XPS are different from those for the OPS.
The OPS active periods appear to precede the
XPS activity.
In contrast, the time scale of the SD
variation seems to be identical in both the
XPS and OPS, at about 100 s.
We checked that the optical light
curves have similar features to the X-ray
curves.

\section{Discussion}

\subsection{Cross-correlation function of XPS and OPS}

\begin{figure}
 \begin{center}
  \includegraphics[width=80mm]{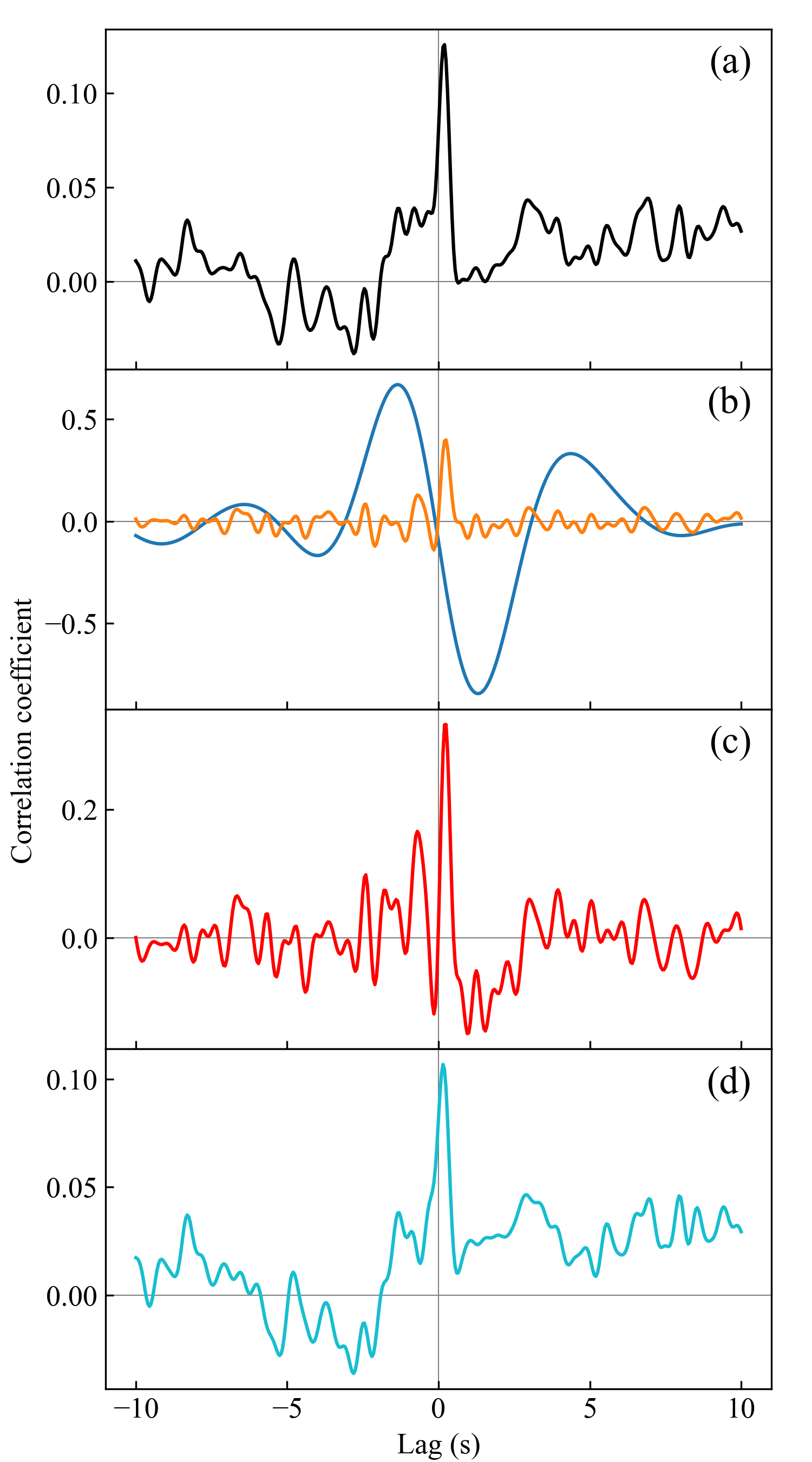}
 \end{center}
 \caption{%
 (a) CCF for the filtered light curves.
 (b) CCFs for the reconstructed light
     curves for the XPS (orange) and OPS (blue).
 (c) CCF for the combined light curves for the
     XPS and OPS.
 (d) CCF for the residual light curves.}%
 \label{fig:ccf}
\end{figure}

The CCF for the filtered light curves is
shown in panel (a) of figure \ref{fig:ccf},
and the CCFs for the XPS and OPS are shown in
panel (b).
The time lag is defined as the optical delay
with respect to the X-ray variation.
The CCF for the XPS has a single peak at a
lag of 0.1 s.
The CCF for the OPS has a single peak at a lag of
$-1$ s and trough at a lag of 1 s.
These CCF patterns for the XPS and OPS are
expected from the reconstructed light curves
shown in figure \ref{fig:lc_rec}.
In order to see the reliability of the correlation coefficient,
we randomly picked up 100 reconstructed light curves obtained
with the resampling method and evaluated the flucation empirically.
As a result, we see that the 90\% confidence intervals of the
coefficients were less than 0.1.
In addition, the CCF for the OPS also has a
small trough at $-3\;{\rm s}$, which is
consistent with the anti-correlated signal 
reported in G08.
Adding the reconstructed light curves for
the XPS to those for the OPS, we made combined
light curves for the X-ray and optical emissions.
The CCF for the combined light curves is shown
in panel (c) of figure \ref{fig:ccf}.
The CCF reproduces the major features of
the filtered curve, that is, the peak at
$+0.1$ s and the trough at $+1$ s.
In addition, the original CCF has relatively
high correlation coefficients between $-2$ and
$0$ s, which is reproduced in panel (c) due to
the contribution of the OPS.
Thus, we conclude that the two positively
correlated signal components, that is, the XPS
and OPS, made the original CCF, including the
apparently anti-correlated patterns.

\citet{Gandhi_2010} performed a cross-spectrum
analysis for the data.
Their time lag spectrum shows optical preceding
components between $\sim 0.8$ and $0.3$ Hz
without high coherence.
The components presumably correspond to the OPS
that we found.
Hence, the result of their cross-spectrum
analysis supports the presence of the OPS.

A possible discrepancy between the CCFs is the
correlation coefficient for the lag of $>3$ s;
it is significantly positive in panel (a),
whereas it is almost zero in panel (c).
This implies that the observed light curves
include variations with a longer time scale
and time lag than those discussed here.
The systematically high correlation coefficients
may cause the relatively shallow trough at
$+1$ s in panel (a) compared with that
in panel (c).
The anti-correlation with a lag of $-3$ s may
be reproduced by the OPS, as shown in panel (b),
whereas the noise in the CCF for the XPS
disturbs this feature in the combined CCF.
After constructing the light curves from the
common components, we subtracted them from the
original filtered light curves and made the
residual light curves of the X-ray and optical
emissions.
The CCF for the residual light curves is shown
in panel (d).
The trough at $+1$ s disappears in panel (d)
because the OPS is subtracted.
We can see that the negative correlation at
$-3$ s does not completely disappear in panel
(d), nor does the XPS peak at $+0.15$ s.
These residual correlations imply that the
amplitudes of the XPS and OPS are underestimated
in our analysis.
A part of the common signals may remains in the residual
light curves even after our analysis.

\subsection{Physical interpretation of OPS}

So far, it has been believed that the observed
X-ray and optical light curves include
anti-correlated variations, as indicated by
the CCF.
The complex shape of the CCF can be interpreted
with two components, a broad anti-correlation
component at a lag of $\sim 0$ s and a sharp
positive correlation peak at a lag of a few
subseconds (\cite{gandhi2008rapid},
\cite{durant2011high}).
There are two major explanations for the
anti-correlation (e.g., \cite{uttley2014multi},
and references therein).
One of them appeared in
\citet{veledina2011synchrotron}, where they
explained that the optical emission is the
synchrotron emission from a hot accretion flow
and the Comptonization of the synchrotron
radiation that is dominant in the X-ray regime.
According to their scenario, the optical decay
occurs because of the synchrotron
self-absorption effect
(\cite{veledina2017expanding}).
The other major explanation appeared in
\citet{malzac2004jet}, where they explained
that the X-ray and optical emission share the
common energy reservoir, which leads to their
anti-correlation.
According to these explanations, the X-ray
emission comes from the corona above the
accretion disk, and the optical emission
comes from the jet.
The X-ray flux increases when the corona
receives much energy from the reservoir.
The optical flux increases after the accreting
gas drops into the inner region, and transfers
energy to the jet.

Our results suggest that there is no
anti-correlation but a positive correlation
with an optical lag of $\sim 0.1$ s (XPS) and
a positive correlation with an X-ray lag of
$\sim 1$ s (OPS).
\citet{gandhi2017elevation} found that the
optical variation in V404 Cyg was delayed
with respect to the X-ray variation by
$\sim 0.1$ s.
Their explanation was that its optical emission
originates from the jet, and the optical time
lag corresponds to the electron travel time
from the base of the jet to the optical emission
region.
The XPS in GX $339-4$ has characteristic
features in common with the short-term
variation in V404 Cyg, and they probably have
the same nature.
Here, we consider different physical models of the OPS.
The asynchronous nature of the active periods of XPS and OPS,
as shown in figure \ref{fig:lcNstd}, indicates that the source
of the OPS is distinct from that of the XPS.

First, we consider a typical situation for the
hard state: the optical emission is the
thermal radiation from the standard disk,
and the X-ray emission is from ADAF.
If the time scale of 5 s corresponds to the
dynamical time scale of the standard disk,
$t_{\rm dyn} = 2\pi r^{2/3} {\sqrt{GM}}$,
the distance between the optical source and
the central BH is $r\sim10^3R_{\rm s}$
(where $R_{\rm s} = GM/c^2$) for a BH mass of
$M=6M_\odot$ \citep{hynes2003dynamical}.
This distance provides an upper limit on the
size of ADAF.
The time lag of 1 s may be explained by the
inward free fall of the optical source in ADAF.
The truncated radius of the disk is
theoretically predicted to be between
$10^2R_{\rm s}$ and $10^4R_{\rm s}$
\citep{esin1997advection},
which is consistent with the above scenario.
However, it is difficult to reproduce the
observed amplitude of the OPS with such an inner,
small emitting source because the thermal
emission from the outer disk ($\sim 10^5
R_{\rm s}$) dominates the optical emission.

Second, we consider a situation in which the
optical emission is synchrotron radiation
from the inner accretion flow.
Strong synchrotron emissions are expected in
magnetically dominated accretion flow (MDAF),
in which electrons can be accelerated at the
site of the magnetic reconnection 
(\cite{dal2010role},
\cite{khiali2015magnetic}).
In this scenario, the origin of the X-ray
time lag is uncertain.
Possibly, the X-ray emission is generated by
the inverse-Compton scattering of synchrotron
emission at the corona, which is distant
from the optical source.
The location of the optical source can be
estimated, in the same way as above, to be
$\sim 10^3R_{\rm s}$.
The time lag of 1 s corresponds to $\sim
10^4R_{\rm s}$ if we consider it as the
light travel time.
However, both distances are much larger
than those expected for the sizes of MDAF
and hot corona ($\lesssim 10R_{\rm s}$).

\citet{kalamkar2016detection} investigate the correlation between
X-ray and infrared light curves and found X-ray lagged signal.
\citet{malzac2018jet} interpreted this as a result of internal
shocks generated by the collisions of plasma shells.
In this model, the infrared flux depends on the difference of
the Lorentz factors of the colliding shells.
Hence, on long time-scales, it corresponds to the time derivative
of the jet Lorentz factor which can be traced by the X-ray flux.
It leads to a positive correlation with $-\pi/2$ phase shift of the
infrared signal to the X-ray one.
According to this scenario, we can also expect a negative
correlation with a $+\pi/2$ phase shift when we consider a
periodic signal as we know from the CCF for OPS.
The two strong positive/negative correlations seen in the CCF of
OPS in panel (b) of figure \ref{fig:ccf} may support the scenario.

\section{Summary}
\label{sec:summary}

This paper describes a study of the signal
propagation between two light curves of
different wavelengths.
More precisely, common signal components were
extracted from the X-ray and optical light
curves of GX $339-4$.
Our method is based on the short time Fourier
analysis, and it assumes the sparsity in the
power spectrum to effectively extract
common signal components from noisy data.
We detected two common signal components:
one is that correlates with a 0.15-s optical
lag and the other is that anti-correlates with a
1-s optical lag.
We found that the reconstructed light curve
for the anti-correlated component showed
a positive correlation with a 1-s X-ray
time lag.
Hence, our explanation is that the data contain
two positive correlated signal components, that
is, the XPS and OPS.
Our results demonstrate that combining the XPS
and OPS reproduces the CCF for the data.
The short time scale and large variation
amplitude of the OPS suggest that the optical
source of the OPS is not the thermal disk, but
rather the inner MDAF or in a jet.

\section*{Acknowledgments}

The authors appreciate Prof. Poshak Gandhi, who
kindly shared data with us and gave useful
comments on this paper.

\appendix

\section{Significance test for the concentration of the Fourier components}\label{ap:signiftest}

\begin{figure*}
\begin{minipage}{0.5\hsize}
 \begin{center}
  \includegraphics[width=80mm]{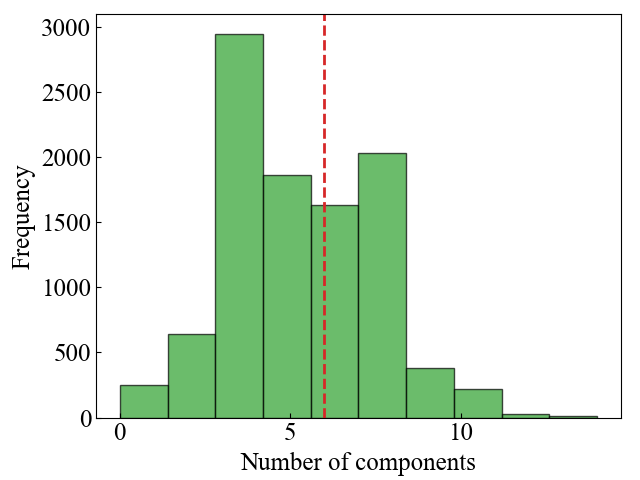}
 \end{center}
\end{minipage}
\begin{minipage}{0.5\hsize}
 \begin{center}
  \includegraphics[width=80mm]{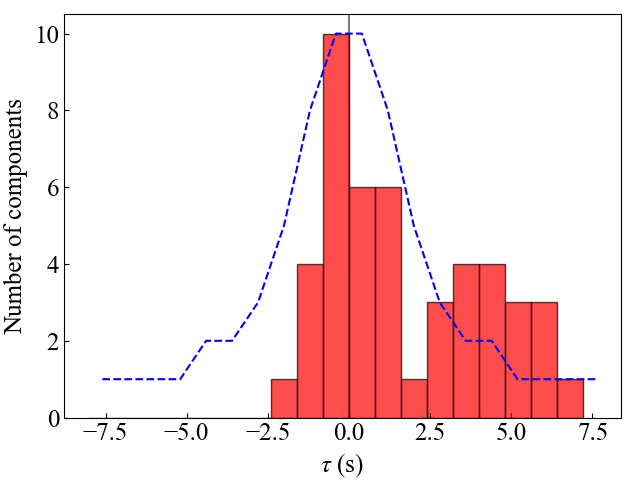}
 \end{center}
\end{minipage}
\caption{%
 Examples of the significance test.
 The data are the same as the simulated data
 used in section \ref{ssec:cs}, which have a
 true lag of $\tau=5$ s.
 Left: The resampling distribution of the
 number of components between $\tau=1$ to 2 s.
 The dashed red line indicates the observed
 number of components.
 Right: The histogram of components, which
 is the same as the red bins in the left
 panel of figure \ref{fig:cbp_art}.
 The blue curve indicates the $90\%$ confidence
 level.
 }%
\label{fig:resampling}
\end{figure*}

As mentioned in section \ref{ssec:cs}, the CS
analysis makes a spurious concentration of the
high-frequency components around $\tau=0$ s.
We employed a significance test to extract the
Fourier components of a common signal with
$\tau \sim 0$ s.
The null hypothesis of the test is that the
Fourier components selected by the CS analysis
provide a totally uncorrelated signal.
In other words, the phase lag of each component
should be uniformly random between $-\pi$ and $\pi$.
We can obtain the distribution of the number of
components in each time lag bin by resampling
the phase lag of the components.
Using the simulated data that are described
in section \ref{ssec:cs}, we show the
resampling distribution of the number of
components between $\tau=1$ and $2$ s in the
left panel of figure \ref{fig:resampling}.
The observed number of components represented
by the red line is almost at the peak of
the distribution, which means that we cannot
reject the null hypothesis.
In this case, we can reject the null-hypothesis
with a $95\,\%$ confidence level if the number
of components is larger than 14.
We show a histogram of the components and the
$95\,\%$ confidence level for each lag bin with
the dashed curve in the right panel of figure
\ref{fig:resampling}.
We can conclude that the concentration of
components at $\tau \sim 5$ s is significant
at a $90\,\%$ of confidence level.
We note that the test is unreliable when
the number of components in a time lag bin
is small.

\section{Relation between $r$ and CCF}\label{ap:r_vs_CCF}

\begin{figure*}
 \begin{center}
  \includegraphics[width=160mm]{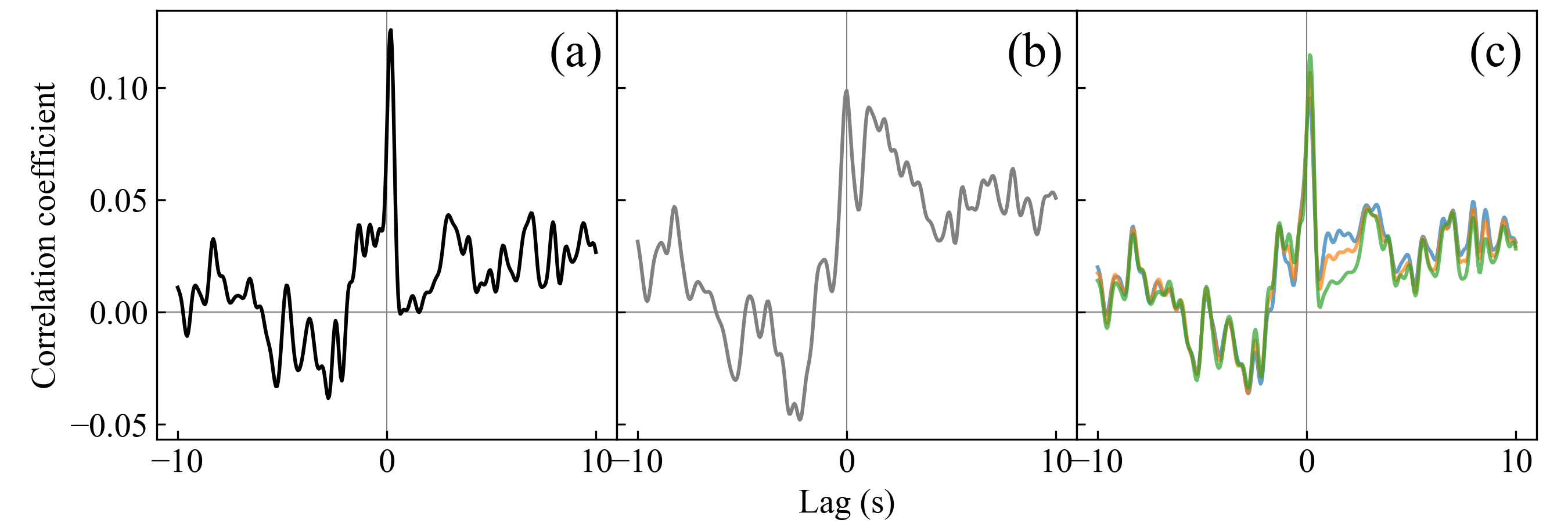}
 \end{center}
 \caption{%
 (a) CCF for the filtered light curves.
 (b) CCFs for the reconstructed light curves which $r=0.0$.
 (c) CCF for the reconstructed light curves which $r=0.7$,
     $0.8$, $0.9$, indicated by blue, orange, and
     green, respectively.}%
 \label{fig:resccf}
\end{figure*}

In the CS analysis, the reconstructed light curves depend on the
parameter of the amplitude ratio, $r$.
In the main text, we used $r=0.8$.
Here, we discuss the dependency of the results on $r$.
The CCF for the filtered light curves is shown in panel (a) of
figure \ref{fig:resccf}.
The CCF for the residual light curves obtained with $r=0.0$ is
shown in panel (b).
The residual CCFs with $r=0.7,\,0.8,\,0.9$ are shown in panel(c)
with  blue, orange, and green curves, respectively.
The orange curve is the same as the CCF in panel (d) of figure
\ref{fig:ccf}.
While we can see a dip around a lag of 1.0 s in the CCF of panel
(a), the CCF rather has a positive peak there in panel (b).
In panel (c), the dip is correctly removed in all residual CCFs.
These results suggest that the reconstructed light curves with
small $r$, such as $r=0.0$, contain a lot of non-common components
which cause an over-fitting for the common signal with a lag of
$\sim 1.0$ s.
Hence, $r$ should be large enough to remove such non-common
components.
The dependence of the results on $r$ is small at $r>0.7$.

\end{document}